\documentclass{iopart}

\usepackage{graphicx}
\usepackage{dcolumn}
\usepackage{bm}
\usepackage{iopams}
\usepackage{multirow}

\begin{document}

\newcommand{\ground}{$5{\rm s}^{2}\,^1{\rm S}_0$~}
\newcommand{\intermed}{$5{\rm s}5{\rm p}\,^1{\rm P}_1$~}
\newcommand{\probe}{$5{\rm s}^{2}\,^1{\rm S}_0 \rightarrow 5{\rm s}5{\rm p}\,^1{\rm P}_1$~}

\title[Spectroscopy of strontium Rydberg states using EIT]{Spectroscopy of strontium Rydberg states using electromagnetically induced transparency}

\author{S Mauger, J Millen and M P A Jones}

\address{Department of Physics, Durham University, Rochester Building, South Road, Durham DH1 3LE, United Kingdom}\ead{m.p.a.jones@durham.ac.uk}

\begin{abstract} We report on the all-optical detection of Rydberg states in an effusive atomic beam of strontium atoms using electromagnetically induced transparency (EIT). Using narrow-linewidth CW lasers we obtain an EIT linewidth of 5 MHz. To illustrate the high spectroscopic resolution offered by this method, we have measured isotope shifts of the $5{\rm s}18{\rm d}\,^1{\rm D}_2$ and $5{\rm s}19{\rm s}\,^1{\rm S}_0$ Rydberg states. This technique could be applied  to  high-resolution, non-destructive measurements of ultra-cold Rydberg gases and plasmas. \end{abstract}

\submitto{\JPB}
\maketitle

Laser-cooled atoms can be excited into high-lying electronic states to  form a
``frozen'' gas of Rydberg atoms~\cite{mourachko98}, where the energy scale associated with the strong dipole-dipole  interactions between the
atoms is much larger than their thermal energy. This system exhibits rich many-body quantum behaviour, such as the ``dipole blockade'' effect~\cite{singer04,tong04,raithel05,vogt06}, where a single Rydberg excitation can be shared by several entangled atoms~\cite{heidemann07}. This effect could be used to realise fast two-qubit quantum logic gates between trapped neutral atoms~\cite{lukin01}.  By  
exciting the atoms above the ionization threshold, an ultra-cold plasma can be created~\cite{killian99}.  These plasmas can reach a strongly coupled regime, where the Coulomb interaction between the ions dominates their kinetic energy, and spatial correlations become important~\cite{killian06}. Rydberg atoms form in these plasmas due to three-body recombination~\cite{killian01}, and spectroscopy of Rydberg atoms inside the plasma could be used to obtain information on the electric field~\cite{dutta00}.

Atoms with two valence electrons such as strontium offer several advantages for experiments on ultra-cold Rydberg gases and
plasmas. Narrow intercombination lines allow laser cooling to temperatures below $1\,\mu$K \cite{katori99}, and extremely high spectroscopic resolution~\cite{boyd06}. In addition, their singly-charged ions have a strong
optical transition in the visible, unlike those of alkali metals which have a closed-shell electronic structure. This allows
the ions in the plasma to be imaged, yielding spatially resolved information on the ion density and temperature~\cite{simien04,chen04}. 

In most experiments on ultra-cold Rydberg gases the Rydberg atoms are detected indirectly by using field
ionization and subsequent detection of the electrons and ions~\cite{gallagher}. This technique is efficient and widely applicable, but is
destructive. Recently~\cite{mohapatra07} , it was shown that electromagnetically induced transparency (EIT) can be used to probe Rydberg
states of rubidium atoms in a vapour cell non-destructively and with high resolution. In combination with strong Rydberg-Rydberg interactions, this could also be used to realise a phase gate between single photons~\cite{friedler05}. 

In this paper we extend this technique to Rydberg states of strontium atoms in a effusive atomic beam. This technique could be used to make high-resolution, real time measurements of the properties of ultracold Rydberg gases and plasmas.
Although EIT
was first observed in strontium~\cite{harris91}, most subsequent work has focussed on the alkali metals. Recently, an ultra-precise optical lattice clock  based on EIT with bosonic $^{88}$Sr atoms was  proposed~\cite{santra05}. This work is the first observation of EIT in strontium using narrow-linewidth CW lasers.

\begin{figure}
\centering
\includegraphics{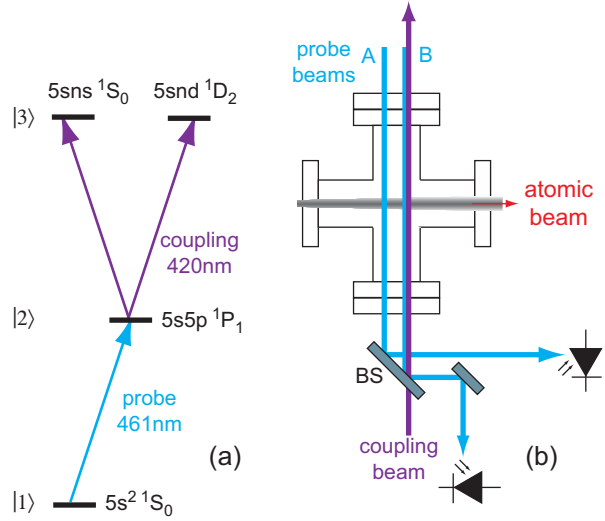}
\caption{(a) Energy level diagram showing the two ladder systems available. (b) Schematic of the experiment. Probe and
coupling beams overlap on a glass plate (BS) and counter-propagate at right angles to the atomic beam.
Two spatially separate probe beams allow simultaneous measurement of the probe beam transmission with and without the
coupling light.\label{fig1}}
\end{figure}

The ladder system studied in these experiments and the experimental setup are shown in figure~\ref{fig1}. The probe beam
is resonant with the  \probe transition at 460.7\,nm, and the coupling beam at 420\,nm
is resonant with transitions from the  \intermed intermediate state to a $5{\rm s}n{\rm s}\,^1{\rm S}_0$ or $5{\rm s}n{\rm d}\,^1{\rm D}_2$ Rydberg
state. Both the probe and the coupling beams are generated using commercial frequency-doubled diode laser systems, and are tuned using a
commerical wavemeter and data on the Rydberg energy levels taken from~\cite{beigang82}. The probe and coupling beams are
overlapped and counter-propagate perpendicularly to the effusive atomic beam of strontium atoms.  A photodiode monitors the transmission of the probe beam as its frequency is scanned across the
transition. A second probe laser beam allows real-time measurement of the probe transmission without the coupling laser
present. Each probe laser beam has a power of 150 $\mu$W and a waist of 1 mm ($1/e^2$ radius), and the two beams are separated by 4 mm. The coupling beam power is typically 12 mW. The probe laser beams are linearly polarized in the vertical direction, and the coupling laser is circularly polarized. This combination was chosen empirically as giving the largest EIT signals under our experimental conditions. To produce the atomic beam, strontium metal is heated in an oven, and the  beam is collimated by a bundle of stainless steel capillaries that restrict the geometric divergence to 43 mrad full width.

\begin{figure}
\centering
\includegraphics{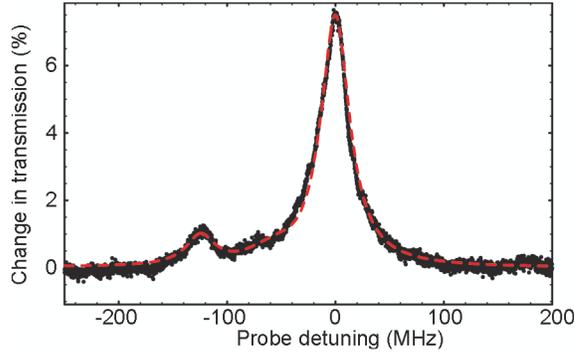}
\caption{Saturated absorption spectrum of the \probe transition (dots). The Doppler background is removed using the second probe beam. The red dashed line shows the fit used to obtain the frequency axis scaling. \label{fig2}}
\end{figure}

To calibrate the probe laser frequency scan, we use saturated absorption spectroscopy and the measured isotope shifts of
the  \probe transition. The saturated absorption spectrum is obtained by replacing the coupling beam with a pump beam at 461\,nm derived from the probe laser.  An example spectrum is shown in figure~\ref{fig2}. The spectrum is
fitted using the sum of six Lorentzian components (one for each isotope and hyperfine component), with the shift  and weighting of
each component taken from table~\ref{table1}, and the width constrained to the natural linewidth of the transition ($\gamma_2=2 \pi \times 16$ MHz). The free parameters are a linear frequency axis scaling and offset, and the peak absorption. The scaling factor and offset are used to calibrate the frequency axis of EIT data  obtained with the same probe laser scan. The accuracy of the scaling factor determines the accuracy of frequency intervals obtained from the EIT data. 
By examining the residuals from this fit we estimate that the uncertainty in the scaling factor is 3\,\%.

\Table{\label{table1} Properties of naturally occuring strontium used to model the spectra. Isotope shifts and hyperfine splittings
for the \probe transition are extracted from \cite{eliel83,kluge74} and are quoted relative to the
principal isotope $^{88}$Sr.}
\br
Isotope & Abundance & $I$ & $F$ & Shift & Relative \\
 & (\%) & & & (MHz)  & strength \\
\mr$^{84}$Sr & 0.56 & 0 & - &  -270.8 & 1 \\
\ms $^{86}$Sr & 9.86 & 0 & - &  -124.5 & 1 \\
\ms \multirow{3}{*}{$^{87}$Sr } &\multirow{3}{*}{7.00} & \multirow{3}{*}{9/2 } & 7/2 & -9.7 & 4/15\\
\ms & & &  9/2 & -68.9 & 1/3 \\
\ms & & & 11/2 & -51.9 & 2/5 \\
\ms $^{88}$Sr & 82.58 & 0 & - &  0 & 1 \\
\br
\endtab

\begin{figure}
\centering
\includegraphics{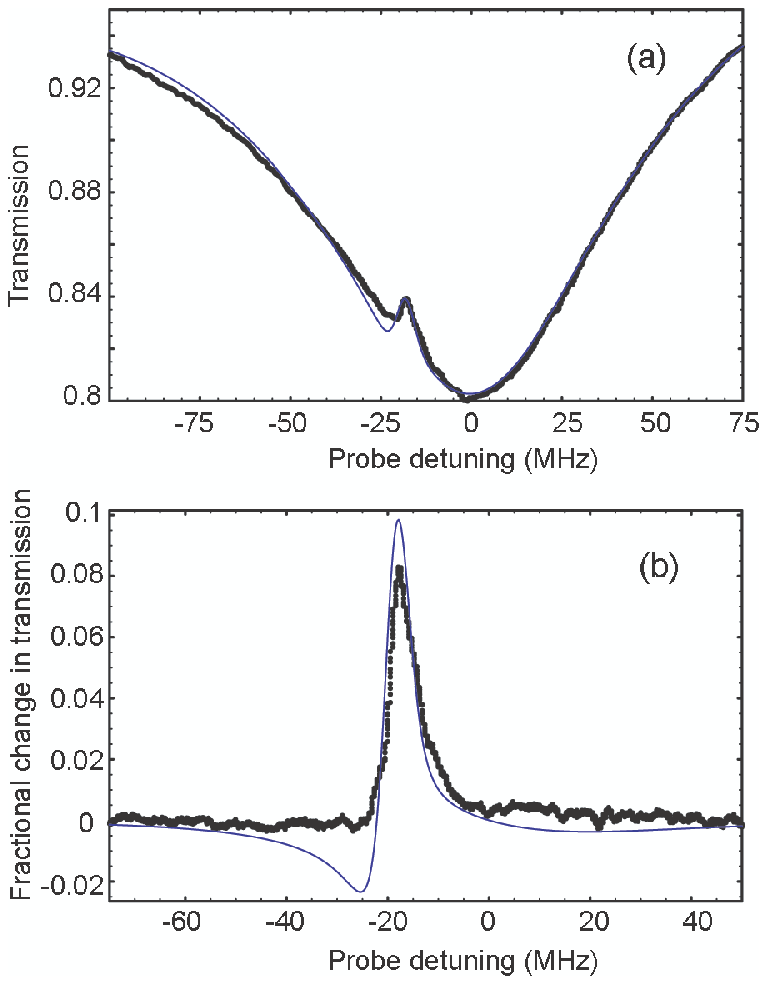}
\caption{(a) Probe transmission as the laser is scanned across the  \probe transition, with
the coupling laser tuned almost on resonance with the $5{\rm s}5{\rm p}\,^1{\rm P}_1 \rightarrow 5{\rm s}18{\rm d}\,^1{\rm D}_2$ transition (dots). Also shown is a fit using equation \ref{eqn1}  with $\gamma_3 = 2 \pi \times 3.5$ MHz, $\Omega_{\rm c} = 2 \pi \times 7.5$ MHz, $\Delta_{\rm c} = 2 \pi \times 20 $ MHz and $\Delta v = 16.5$ ms$^{-1}$ (thin blue line). (b) Difference in transmission between the
two probe beams for the same scan (dots). The thin blue line is the difference between the fit in (a)
and the same curve with $\Omega_{\rm c} = 0$  (all other parameters as in (a)).\label{fig3}}
\end{figure}

If the coupling laser is tuned close to the $5{\rm s}5{\rm p}\,^1{\rm P}_1 \rightarrow 5{\rm s}18{\rm d}\,^1{\rm D}_2$ transition in the dominant $^{88}$Sr isotope,
then we obtain the spectrum shown in figure~\ref{fig3}a. A narrow EIT resonance feature enhances the transmission in a narrow spectral window within the Doppler-broadened absorption profile. The position and amplitude of the resonance depends on the detuning of the
coupling laser. The difference between the transmission of the two probe beams is shown in figure~\ref{fig3}b. The width of the EIT peak is
5\,MHz (FWHM), which is considerably narrower than the natural width of the intermediate  $^1P_1$  state ($\gamma_2=2 \pi \times 16$ MHz), and the maximum
reduction in absorption is 8\%.  With the coupling laser tuned exactly on resonance, we observe a change in absorption
of up to 12\%. 

In order to understand these data in more detail, we have compared them to a model based on an approximate expression for
the susceptibility $\chi(v)$ derived in the limit of a weak probe beam~\cite{xiao95}
\begin{eqnarray}
\fl \chi(v)dv =  -\rmi \frac{3\lambda_{\rm p}^2}{4\pi}\gamma_2 N(v)dv \left[  \gamma_2 - \rmi \left( \Delta_{\rm p} - \bm{k}_{\rm p}\cdot \bm{v} \right)
\right. \nonumber \\
 + \left. \frac{\left( \Omega_c/2 \right)^2}{\gamma_3 - \rmi \left[ \Delta_{\rm p} + \Delta_{\rm c} - \left(\bm{k}_{\rm p} + \bm{k}_{\rm c} \right)
\cdot \bm{v} \right] } \right]^{-1}\label{eqn1} ,
\end{eqnarray}
where $N(v)$ is the number density of atoms with transverse velocity $v$, $\Delta_{\rm p,c}$ and $\bm{k}_{\rm p,c}$ are the
probe and coupling laser detuning and wavevectors respectively, and $\Omega_{\rm c}$ is the coupling laser Rabi frequency. The decay rate $\gamma_2=2 \pi \times 16$ MHz is the natural width of the intermediate state, and $\gamma_3$ is the decay rate of the Rydberg state. This includes the lifetime of the Rydberg state as well as other line-broadening mechanisms, and is treated as a fit parameter.

 The
lineshape is obtained by summing the contribution to $\chi(v)$ from all four isotopes and integrating the imaginary part of
(\ref{eqn1}) over the transverse velocity distribution of the atomic beam. In figure~\ref{fig3}b, the coupling laser is tuned such that only the $^{88}$Sr isotope
 contributes to the EIT resonance part of the signal, and so we set $\Omega_\mathrm{c}=0$  for the other isotopes to simplify the numerical
integration. 

A standard treatment of the transverse velocity distribution of an effusive atomic beam~\cite{demtroder} assumes that the velocity distribution inside the oven follows a Maxwell-Boltzmann distribution, and that the divergence angle $\epsilon$ of the beam is fixed geometrically by  an aperture some distance from the exit slit of the oven. Collisions with the walls of the exit slit or the collimation aperture are neglected. The result is a Gaussian transverse velocity distribution, with a Doppler width that is reduced by a factor
$\sin \epsilon$ from that obtained in a vapour at thermal equilibrium at the same temperature.  
We find that using  a Gaussian velocity distribution, we are unable to describe the shape of the Doppler broadened
absorption profile. This is almost certainly due to collisions with the walls of the capillaries (length=8mm, diameter = 0.17 mm) used to collimate the atomic beam, which modify the transverse velocity distribution. Empirically,  we find that a Lorentzian distribution $N(v) \propto n_{\rm at} / (1+ v^2/\Delta v^2)$ for the transverse velocity  gives
good agreement with our data. As a result, there are five adjustable parameters in our model: the
half-width of the Lorentzian velocity distribution $\Delta v$, the density of the atomic beam $n_{\rm at}$, the parameters $\Delta_{\rm c}$, $\Omega_{\rm c}$ of the coupling laser and the rate $\gamma_3$. 

The results of this model are compared with our data in figures \ref{fig3}a and \ref{fig3}b. The model accurately
reproduces the shape of the Doppler background in figure~\ref{fig3}a, and the asymmetry of the EIT peak in  figure
\ref{fig3}b. By looking at the behaviour of the model in more detail, we see that the latter effect arises from the
combination of the wavelength mismatch between the probe and coupling beams and the velocity averaging. However, we do
not observe the {\it increase} in absorption that is predicted to occur on both sides of the EIT peak, and which was observed
in similar experiments in rubidium~\cite{mohapatra07}. 
\begin{figure}
\centering
\includegraphics{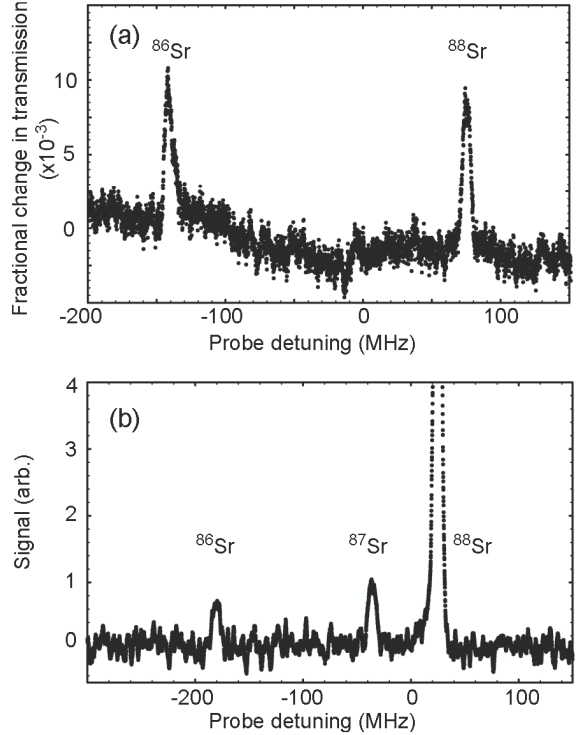}
\caption{(a) Probe beam intensity difference with the coupling laser tuned slightly away from resonance with the $5{\rm s}5{\rm p}\,^1{\rm P}_1 \rightarrow 5{\rm s}18{\rm d}\,^1{\rm D}_2$  transition. A second 
EIT peak appears due to the $^{86}$Sr isotope. (b) EIT signal with the coupling laser tuned near the  $5{\rm s}5{\rm p}\,^1{\rm P}_1 \rightarrow 5{\rm s}19{\rm s}\,^1{\rm S}_0$ transition. Using lock-in detection, all three abundant isotopes can be observed. The peak due to the  $^{88}$Sr isotope is visible without the lock-in and gives a $\sim5\,\%$
 change in absorption. 
\label{fig4}}
\end{figure}

By varying the coupling laser detuning, we are also able to observe EIT signals due to the other isotopes. Figure~\ref{fig4}a shows
EIT signals due to both the $^{86}$Sr and  $^{88}$Sr isotopes. We do not observe a signal due to the $^{87}$Sr
isotope. The hyperfine interaction splits the $^{87}$Sr line into nine components, and for $15 \le n \le 25$ the 
hyperfine splitting of the $5{\rm s}n{\rm d}\,^1{\rm D}_2$ Rydberg states is very large ($\sim 1$ GHz) due to strong mixing with nearby triplet states~\cite{beigang81, beigang82b}. As a result of this it is difficult to observe an EIT signal due to this isotope. In contrast, no hyperfine splitting occurs for the $5{\rm s}n{\rm s}\,^1{\rm S}_0$ states. 
The EIT signal that we obtain with the coupling laser tuned to the $5{\rm s}5{\rm p}\,^1{\rm P}_1 \rightarrow 5{\rm s}19{\rm s}\,^1{\rm S}_0$ transition is
shown in figure~\ref{fig4}b. In this experiment, the amplitude of the coupling laser was modulated using a chopper, and with lock-in detection we observed the signal from the $^{86}$Sr, $^{87}$Sr and $^{88}$Sr isotopes. Due to the wavelength mismatch between the pump and probe lasers and the velocity averaging, the hyperfine splitting of the \intermed intermediate state is scaled by a factor $1-\lambda_{\rm c} / \lambda_{\rm p}$. This results in a separation between the EIT signals for each hyperfine component of less than 5 MHz, which is not resolved in figure~\ref{fig4}b.

The frequency intervals measured in  figure~\ref{fig3} and the isotope shifts given in table \ref{table1}  can be used to obtain the isotope shifts of the Rydberg states relative to the \ground  ground state. In addition to the scaling in the intermediate state, the wavelength mismatch also results in frequency separations in the Rydberg state being scaled by  $\lambda_{\rm c} /\lambda_{\rm p}$. The results are given in table \ref{table2}. Our measurement of the  $^{88}$Sr - $^{86}$Sr isotope shift on the $5{\rm s}^2\,^1{\rm S}_0 \rightarrow 5{\rm s}18{\rm d}\,^1{\rm D}_2$  transition is in agreement with a previous value of $229\pm5$ MHz reported in~\cite{lorenzen83}. The uncertainty is dominated by the uncertainty in the frequency axis calibration. This could be substantially improved by scanning the laser with an acousto-optic modulator, or by using a well-characterised etalon to calibrate the scan. The narrow linewidth of the EIT features would then allow isotope shifts and hyperfine splitting to be determined with better than 1 MHz resolution.

\Table{\label{table2} Transition isotope shifts between the Rydberg states and the ground state calculated from the intervals measured in figure~\ref{fig3} and the data in table \ref{table1}.}
\br
Transition & $^{88}$Sr - $^{86}$Sr & $^{88}$Sr - $^{87}$Sr  \\
   & (MHz) & (MHz)   \\
\mr
  $5{\rm s}^2\,^1{\rm S}_0 \rightarrow 5{\rm s}18{\rm d}\,^1{\rm D}_2$ & $226\, \pm \,7$ & --\\
\ms $5{\rm s}^2\,^1{\rm S}_0 \rightarrow 5{\rm s}19{\rm s}\,^1{\rm S}_0$ & $213\, \pm \,7$ & $62 \, \pm \,8$\\
\br
\endtab

In summary we have shown that electromagnetically induced transparency can be used to probe the Rydberg states of  strontium atoms with high resolution.    As an illustration, isotope shifts were measured for transitions from the ground state to the $5{\rm s}18{\rm d}\,^1{\rm D}_2$ and  $5{\rm s}19{\rm s}\,^1{\rm S}_0$ states. This technique could have important applications as a non-destructive, high-resolution probe of strongly interacting Rydberg gases and plasmas.
  \begin{ack} 
We are grateful to C.~S.~Adams, I.~G.~Hughes, A.~Mohapatra and A. Browaeys for many useful discussions. We also thank C.~S.~Adams for the loan of equipment and the EPSRC for financial support.
 \end{ack}
\bibliographystyle{unsrt.bst}
\bibliography{strontium_eit_arXiv}
\end{document}